\documentclass[ aps, showpacs, showkeys, nofootinbib, floatfix, superscriptaddress]{revtex4}

\usepackage{amsfonts}

\usepackage{amssymb}
\usepackage{amsmath}
\usepackage{graphicx}

\begin{document}

\title{Linearized modified gravity theories and gravitational waves physics in the GBD  theory}
 \author{Jianbo Lu}
 \email{lvjianbo819@163.com}
 \affiliation{Department of Physics, Liaoning Normal University, Dalian 116029, P. R. China}
 \author{Yan Wang}
 \affiliation{Department of Physics, Liaoning Normal University, Dalian 116029, P. R. China}
\author{Xin Zhao}
 \affiliation{Department of Physics, Liaoning Normal University, Dalian 116029, P. R. China}

\begin{abstract}
 The  generalized Brans-Dicke (abbreviated as GBD) theory is obtained by replacing the Ricci scalar $R$  in the original Brans-Dicke (BD) action with an arbitrary function $f(R)$.  Comparing with other theories,  some interesting properties have been found or  some problems existing in other theories could be solved in the GBD theory. For example, (1) the state parameter of geometrical dark energy in the GBD model can cross over the phantom boundary $w=-1$,  without bearing the problems existing in the quintom model (e.g. the problem of negative kinetic term and the fine-tuning problem, etc); (2)  $f(R)$ theory is equivalent to the BD  theory with a potential (BDV) with the couple parameter $\omega=0$, where the kinetic term of field in the equivalent BDV theory is absent. While the scalar fields in the GBD own the non-disappeared kinetic term, when one compares the GBD theory with the $f(R)$ theory.  In this paper, we continue to investigate the GBD theory.  Using the method of the weak-field approximation, we explore the linearized physics in the   GBD theory. The linearized equations of the gravitational field and two scalar fields  are given.  We investigate their solutions in the linearized theory for a point mass. It is shown that the problem of the $\gamma$ (parametrized post-Newtonian parameter) value in the $f(R)$ theory could be solved in the GBD theory, where the theoretical $\gamma$ value in the GBD theory can be consistent with the observational results.  At last, we study the gravitational waves physics in the vacuum for the GBD theory. It is found that the gravitational radiation in the GBD theory has new freedoms beyond the two standard modes in the general relativity theory.

\end{abstract}



\maketitle

\section{$\text{Introduction}$}

 The several observations in the 1990s  showed that the  expansion of our universe is accelerating \cite{SN-acc1,SN-acc2}. According to the famous general relativity (GR) theory, this accelerating expansion is caused by the dark energy with the negative pressure. The most popular candidate of dark energy from the viewpoint of the observations is the cosmological constant with the equation of state $w=-1$, though it exists the fine-tuning and the coincidence problems in the theory. Another popular candidate of dark energy model is the quintessence scalar field. The problem for this scalar-field model is that we have to introduce the scalar field and its potential by the hand. Also, other dark energy models \cite{DE,DE1,DE2,DE3,DE5,DE6,DE7,DE9,DE10,DE-lu1,DE-lu2} exist the respective problems.

 GR as the standard model of gravity is tested well, especially in the solar system. But it also confronts many unanswered questions. Several observational and theoretical  motivations require us to investigate the  modified or alternative theories of GR. For examples, it is hard to  combine quantum physics with the principles of GR, or the GR does not lead to a renormalizable model, etc. Studies on the modified gravity theories of GR has been always a hot area. Several modified gravity theories have been widely studied \cite{mg1,mg2,mg3,mg4,mg5,mg6,mg7,mg8,mg9}, especially two simple  modifications to GR: the $f(R)$ theory \cite{fr-review1,fr-review2} and the  Brans-Dicke (BD)  theory  \cite{original-BD}. In the BD  theory,  a scalar field $\phi$ can be introduced naturally by defining $\phi(t)=1/G(t)$. $G$ is the Newton gravity constant. Obviously, BD theory  is a time-variable $G(t)$ gravity theory which can be sustained by the recent observations \cite{VG-MNRAS-2004-dwarf,VG-PRD-2004-white,VG-APJ,VG-PRD-2002-SN,VG-PRL-1996-neutron,vg-prl-constraint}. For observational and theoretical  motivations, several extended versions of the BD theory have been investigated and developed, such as adding a potential term to the original BD theory \cite{BD-potential},   assuming the coupling constant $\omega$ to be variable with respect to time \cite{BD-omegat1,BD-omegat2}, etc.  The applications of these extended BD theories have been investigated widely, such as at the aspects of cosmology \cite{GBD-cosmic1,GBD-cosmic2,GBD-cosmic3}, weak-field approximation \cite{GBD-weak}, observational constraints \cite{GBD-constraint1,GBD-constraint2}, and so on   \cite{BD-widely1,BD-widely2,BD-widely3}.

Recently, a different method is proposed  to modify the BD theory  (abbreviate as GBD)  in Ref. \cite{GBD}  by generalizing the Ricci scalar $R$ to be an arbitrary function $f(R)$ in the BD action.  Comparing with other theories, it can be found   in Ref. \cite{GBD} that the GBD theory have some interesting properties or solve some problems existing in other theories. (1)  One knows that in the couple scalar-fields quintom model, it is required to include both the canonical quintessence field and the non-canonical phantom field  in order to make the state parameter to cross over $w=-1$, while several fundamental problems are associated with phantom field, such as the problem of negative kinetic term and the fine-tuning problem, etc. While, in the GBD model, the state parameter of geometrical dark energy can cross over the phantom boundary $w=-1$ as achieved in the quintom model,  without bearing the problems existing in the quintom model. (2) It is well known that the $f(R)$ theory are equivalent to the BD theory with a potential (abbreviated as BDV) for taking a specific value of the BD parameter $\omega=0$, where the specific choice: $\omega=0$ for the BD parameter is quite exceptional, and it is hard to understand the corresponding absence of the kinetic term for the field. However, for the GBD theory, it is similar to the couple scalar-fields model, and both fields in the GBD own the non-disappeared dynamical effect.  In addition, the GBD theory tends to investigate the physics from the viewpoint of geometry, while the BDV or the two scalar-fields quintom model tends to solve physical problems from the viewpoint of matter. It is possible that several special characteristics of scalar fields could be revealed through studies of geometrical gravity in the GBD. Furthermore, the GBD theory (just as the so-called $f(\phi)R$ theory \cite{fphiR,fphiR1}) can be considered as a special case of the more complex  $f(R,\phi)$ theory \cite{fR-phi,fR-phi1,fR-phi2,fR-phi3}. It is well known that the so-called $f(\phi)R$ theory \cite{fphiR,fphiR1},  as a special case of the $f(R,\phi)$ theory, has been widely studied \cite{fphiR2,fphiR3,fphiR4}. Given that $f(R,\phi)$ is a more complex theory and the more simple theory is usually more favored by the researcher in physics, here we  continue to explore the GBD theory.

 Linearized theory is a weak-field approximation to gravity theory, which  is a good method to test the gravity theories alternative to GR according to the current observations. In fact, this approximation is well applicative in nature except for phenomenon dealing with the large scale structure of the universe and phenomenon dealing with black holes and gravitational collapse. The weak-field approximation method has been studied in lots of gravity theories, such as the higher order gravity \cite{wf-higher-order,wf-higher-order1,wf-higher-order2,wf-higher-order3}, the conformal Weyl gravity \cite{wf-conformal}, the Galilean gravity \cite{wf-galilean}, the infinite derivative gravity \cite{wf-derivative}, the $f(R)$ gravity \cite{wf-fr1,wf-fr2,wf-fr3}, the $f(T)$ theory \cite{wf-ft}, the scalar-tensor gravity \cite{wf-st1,wf-st3}, the Horava-Lifshitz gravity \cite{wf-hl}, etc. In this paper, we discuss the linearized modified gravity theories of GBD. We derive to give the linearized GBD field equations, and  find their solutions for a point mass. In addition, we perform the  calculation on the post Newton parameter, and it is shown that the GBD theory can solve the problem about the  variance between the theoretical value and the observational value of post Newton parameter $\gamma$ in the popular $f(R)$ modified gravity theory.

The detection of gravitational waves (GWs) by the LIGO Collaboration is a milestone in GW research and opens a new window to probe gravity theory and astrophysics \cite{ligo-gws1,ligo-gws2,ligo-gws3}. Future GWs observations will offer more accurate data, so it is worthwhile to investigate GWs physics in alternative theories of gravity. GWs physics have been studied in several modified gravity theories, such as the scalar-tensor theories \cite{gw-st1,gw-st2,gw-st3}, the $f(R)$ theories \cite{gw-fr1,gw-fr2,gw-fr3}, the conformal gravity \cite{gw-cf}, the $f(T)$ theories \cite{gw-ft}, the mimetic gravity \cite{gw-mimetic,gw-mimetic1}, the non-minimal curvature-matter coupling theory \cite{gw-coupling}, etc. In this paper, we study the gravitational radiation in GBD theory. We derive to give the equations of gravitational radiation, which will be valuable  to test gravity theories alternative to GR for future observations of GWs.

This paper is organised as follows. The gravity field equation and the weak-field approximation equations are given in Section II. Solutions to the linearized GBD field equations for a point mass and the calculations on the post Newton parameter are investigated in Section III.  In Section IV, the gravitational waves physics are studied in this part.  Section V is the conclusion and the discussion on our results.

\section{$\text{ Weak field equations in GBD theory}$}

In the framework of time-variable gravitational constant, we study a generalized Brans-Dicke theory by using a function $f(R)$ to replace the Ricci scalar $R$  in the original BD action. The action of system is written as
\begin{equation}
S=\frac{1}{2}\int d^4x{\cal L}_{T}=\frac{1}{2}\int \sqrt{-g}[\phi f(R)- \frac{\omega}{2\phi}\partial _\mu \phi \partial ^\mu \phi+\frac{16\pi }{c^4}L_m]d^4x.\label{action}
\end{equation}
Obviously, the system contains three dynamical variable: the gravitational field $g_{\mu \nu}$, the matter field $\psi$ and the scalar field $\phi$. $\omega$ is the couple constant. Varying the action (\ref{action}), one can get the gravitational field equation and  the BD scalar field equation as follows
\begin{eqnarray}
\phi \left[ f_{R}R_{\mu \nu }-\frac{1}{2}f(R)g_{\mu \nu }\right]- (\nabla_\mu\nabla_\nu -g_{\mu\nu}\Box)(\phi f_{R})+ \frac{1}{2}\frac{\omega}{\phi}g_{\mu\nu}\partial_\sigma\phi\partial^\sigma\phi
-\frac{\omega}{\phi}\partial_\mu\phi\partial_\nu\phi = 8\pi T_{\mu \nu },\label{gravitational-eq}
\end{eqnarray}
\begin{equation}
f(R)+2\omega\frac{\Box \phi}{\phi} -\frac{\omega}{\phi^{2}}\partial _\mu \phi \partial ^\mu \phi=0,\label{scalar-eq}
\end{equation}
where $f_{R}\equiv \partial f/\partial R$, $\nabla _\mu $ is the covariant derivative associated with the Levi-Civita connection of the metric,  $\Box \equiv \nabla ^\mu \nabla _\mu $, and $T_{\mu \nu }=\frac{-2}{\sqrt{-g}}\frac{\delta S_m}{\delta g^{\mu \nu }}$ is the energy momentum tensor of the matter.
The trace of Eq. (\ref{gravitational-eq}) is
\begin{eqnarray}
f_{R}R-2f(R)+\frac{3\Box (\phi f_{R})}{\phi}+\frac{\omega}{\phi^{2}}\partial_\mu\phi\partial^\mu\phi
= \frac{8\pi T}{\phi}.\label{trace}
\end{eqnarray}
Obviously,  the standard $f(R)$ modified gravity is recovered for $\phi$=constant, while the original BD theory is obtained for taking $f(R)=R$ in the above equations. Combining Eqs. (\ref{scalar-eq}) and (\ref{trace}), we get
 \begin{eqnarray}
\Box \phi-\frac{\partial_{\mu}\phi\partial ^{\mu}\phi}{4\phi}=\frac{1}{4\omega}[8\pi T -\phi R f_{R}-3\Box (\phi f_{R})].\label{dynamical-phi}
\end{eqnarray}

Having the correct weak-field limit at the Newtonian and the post-Newtonian levels is a crucial issue that has to be addressed for any viable alternative gravitational  theory of GR. In this  weak-field limit mehtod, it means that the spacetime metric is nearly flat and  can be expanded as
\begin{eqnarray}
g_{\mu\nu}=\eta_{\mu\nu}+h_{\mu\nu},\label{weak-g}
\end{eqnarray}
here $\eta_{\mu\nu}$ is the Minkowski metric, and $h_{\mu\nu}$ denotes a small  deviate with respect to the flat spacetime, i.e. $|h_{\mu\nu}|\ll 1$. The inverse metric is $g^{\mu\nu}=\eta^{\mu\nu}-h^{\mu\nu}$, where the Minkowski metric is used to raised the indices. The trace $h$ is given by $h=\eta^{\mu\nu}h_{\mu\nu}$. According to Eqs.(\ref{trace}) and (\ref{dynamical-phi}), we see that the GBD could be considered as two scalar-fields theory, i.e. the BD field and the effectively geometrical field $f_{R}=\Phi$. For these two scalar fields, the weak-field approximations can be expressed as
 \begin{eqnarray}
\phi=\phi_{0}+\varphi,\label{weak-phi}
\end{eqnarray}
 \begin{eqnarray}
\Phi=\Phi_{0}+\delta\Phi,\label{weak-Phi}
\end{eqnarray}
with $|\varphi|\ll \phi_{0}$ and $|\delta\Phi|\ll \Phi_{0}$.

Ignoring the second-order and the higher terms, we can express the linearized equations of the gravitational field and the both scalar fields as
 \begin{eqnarray}
\tilde{R}_{\mu\nu}-\frac{\tilde{R}}{2}\eta_{\mu\nu}=
\partial_{\mu}\partial_{\nu}\frac{\delta\Phi}{\Phi_{0}}+\partial_{\mu}\partial_{\nu}\frac{\varphi}{\phi_{0}}
-\eta_{\mu\nu}\Box_{\eta}\frac{\delta\Phi}{\Phi_{0}}-\eta_{\mu\nu}\Box_{\eta}\frac{\varphi}{\phi_{0}}+\frac{8\pi T_{\mu\nu}}{\phi_{0}\Phi_{0}},\label{eq-weak-gravity}
\end{eqnarray}
 \begin{eqnarray}
\Box_{\eta}\varphi=\frac{3}{4\omega+3\Phi_{0}}[8\pi T-\phi\tilde{R}\Phi_{0}-3\phi_{0}\Box_{\eta}\delta\Phi],\label{eq-weak-phi}
\end{eqnarray}
 \begin{eqnarray}
\Box_{\eta}\frac{\delta\Phi}{\Phi_{0}}=\frac{2f(\tilde{R})}{3\Phi_{0}}-\frac{\tilde{R}}{3}-\frac{\Box_{\eta}\varphi}{\varphi_{0}}+\frac{8\pi T}{3\phi_{0}\Phi_{0}},\label{eq-weak-Phi}
\end{eqnarray}
with  $\Box_{\eta}=\partial^{\sigma}\partial_{\sigma}$.  Here $\tilde{R}_{\mu\nu}$ and $\tilde{R}$  denote  the linearized Ricci tensor and the linearized Ricci scalar. Defining a new tensor $\theta_{\mu\nu}$ and  choosing a gauge (as called the Lorenz gauge or the Harmonic gauge in GR) as follows
\begin{eqnarray}
\partial^{\nu}\theta_{\mu\nu}=(h_{\mu\nu}-\frac{1}{2}\eta_{\mu\nu}h-\eta_{\mu\nu}\frac{\varphi}{\phi_{0}}+\eta_{\mu\nu}h_{f})^{,\nu}
=h_{\mu\nu}^{,\nu}-\frac{1}{2}h_{,\mu}-\frac{\varphi_{,\mu}}{\phi_{0}}+h_{f,\nu}=0,\label{Lorentz-gauge}
\end{eqnarray}
with $h_{f}\equiv \frac{\delta\Phi}{\Phi_{0}}$, we get expressions of the  linearized Ricci tensor and Ricci scalar as
\begin{eqnarray}
\tilde{R}_{\mu\nu}=\frac{1}{2}(-2\partial_{\mu}\partial\nu h_{f}+2\partial\mu\partial\nu\frac{\varphi}{\phi_{0}}-\Box_{\eta}\theta_{\mu\nu}+\frac{\eta_{\mu\nu}}{2}\Box_{\eta}\theta-\eta_{\mu\nu}\Box_{\eta} h_{f}+\eta_{\mu\nu}\Box_{\eta}\frac{\varphi}{\phi_{0}}),\label{linear-Rmunu}
\end{eqnarray}
\begin{eqnarray}
\tilde{R}=-3\Box_{\eta} h_{f}+3\Box_{\eta}\frac{\varphi}{\phi_{0}}+\frac{\Box_{\eta}\theta}{2}.\label{linear-R}
\end{eqnarray}
Here $\theta=\eta^{\mu\nu}\theta_{\mu\nu}=-h+4h_{f}-4\frac{\varphi}{\phi_{0}}$. Combining Eqs. (\ref{linear-Rmunu},\ref{linear-R}) with Eqs. (\ref{eq-weak-gravity}-\ref{eq-weak-Phi}), to the first order we obtain the linearized gravitational field equation and two linearized scalar-field equations in the GBD theory, respectively, as follows
\begin{eqnarray}
\Box_{\eta}\theta_{\mu\nu}=-\frac{16\pi T_{\mu\nu}}{\phi_{0}\Phi_{0}},\label{eq-box-theta}
\end{eqnarray}
\begin{eqnarray}
\Box_{\eta}\varphi=\frac{8\pi T}{2\omega+3\Phi_{0}},\label{eq-box-varphi}
\end{eqnarray}
\begin{eqnarray}
\Box_{\eta} h_{f}-m_{s}^{2}h_{f}=\frac{16\pi\omega T}{3\phi_{0}\Phi_{0}(2\omega+3\Phi_{0})}.\label{eq-box-Phi}
\end{eqnarray}
Here constant $m_{f}\equiv\frac{\Phi_{0}}{\delta \Phi}(\frac{2f}{3\Phi_{0}}-\frac{\tilde{R}}{3})$  has mass dimension with referring  the discussion in Ref. \cite{weak-fr-mf1,weak-fr-mf2,weak-fr-mf3}. In the following, we solve these three linearized field equations for case of a static point mass. And then we derive to give the theoretical expressions of the parametrized post-Newtonian (PPN) parameter $\gamma$. At last, on the basis of the Eqs. (\ref{eq-box-theta}-\ref{eq-box-Phi}), we consider the  vacuum GWs physics in the GBD theory.

\section{$\text{Solutions to the linearized field equations for a point mass and discussions on PPN parameter in GBD }$}
\subsection{$\text{Solutions to the linearized field equations for a point mass in GBD theory}$}
In this part, we investigate to obtain a physically relevant  solution to the linearized GBD theory. Considering a point mass term as a source, the energy momentum tensor of the point particle is described by
\begin{equation}
T_{\mu\nu}=m\delta(\vec{r})diag(1,0,0,0).\label{Tmunu-point}
\end{equation}
Obviously, here point particle is located at $\vec{r}=0$ with  $\vec{r}^{2}=\vec{x}^{2}+\vec{y}^{2}+\vec{z}^{2}$. Substituting  $T=\eta^{\mu\nu}T_{\mu\nu}=-m\delta(\vec{r})$ into Eq.(\ref{eq-box-varphi}-\ref{eq-box-Phi}) and considering a static state, we obtain the expressions of perturbation variables of the BD scalar field $\phi$ and the geometrical field $\Phi$ respectively
\begin{equation}
\varphi(r)=\frac{2m}{2\omega+3\Phi_{0}}\frac{1}{r},\label{phi-r}
\end{equation}
\begin{equation}
h_{f}(r)=\frac{4\omega m}{3\phi_{0}\Phi_{0}(2\omega+3\Phi_{0})}\frac{e^{-m_{s}r}}{r},\label{hf-r}
\end{equation}
Furthermore, we derive the non-zero expression of the perturbation variable of the metric field in the GBD theory. For the "00" component in Eq. (\ref{eq-box-theta}), we get the solution
\begin{equation}
\theta_{00}=\frac{4m}{\phi_{0}\Phi_{0}}\frac{1}{r}.\label{theta00}
\end{equation}
Using the relation $\theta=-h+4h_{f}-4\frac{\varphi}{\phi_{0}}$, we have
\begin{equation}
h_{\mu\nu}=\theta_{\mu\nu}-\eta_{\mu\nu}\frac{\theta}{2}+\eta_{\mu\nu}h_{f}-\eta_{\mu\nu}\frac{\varphi}{\phi_{0}}.\label{hmunu}
\end{equation}
According to Eqs.(\ref{phi-r}-\ref{hmunu}) and using the relation $\theta=\eta^{\mu\nu}\theta_{\mu\nu}=-\frac{4m}{\phi_{0}\Phi_{0}}\frac{1}{r}$,
we obtain  the non-vanishing components of the metric perturbation term as follows
\begin{equation}
h_{00}=\frac{2m}{\phi_{0}\Phi_{0}r}+\frac{2m}{\phi_{0}(2\omega+3\Phi_{0})r}-\frac{4\omega m}{3\phi_{0}\Phi_{0}(2\omega+3\Phi_{0})}\frac{e^{-m_{s}r}}{r},\label{h00}
\end{equation}
\begin{equation}
h_{ij}=\frac{2m}{\phi_{0}\Phi_{0}r}-\frac{2m}{\phi_{0}(2\omega+3\Phi_{0})r}+\frac{4\omega m}{3\phi_{0}\Phi_{0}(2\omega+3\Phi_{0})}\frac{e^{-m_{s}r}}{r}.\label{hij}
\end{equation}
Here $i,j =1,2,3$ denote the space index, and the Greece letters denote the spacetime index. The last two terms of the right hand in Eqs. (\ref{h00}) and  (\ref{hij}) describe the effects of two scalar fields ($\phi$ and $\Phi$) in the metric perturbation tensor  by relating to the Eqs. (\ref{phi-r}) and (\ref{hf-r}). So, we can lastly receive the expressions of the non-zero field variables as follows
\begin{equation}
g_{00}=-1+\frac{2m}{\phi_{0}\Phi_{0}r}+\frac{2m}{\phi_{0}(2\omega+3\Phi_{0})r}-\frac{4\omega m}{3\phi_{0}\Phi_{0}(2\omega+3\Phi_{0})}\frac{e^{-m_{s}r}}{r},\label{g00}
\end{equation}
\begin{equation}
g_{ij}=[1+\frac{2m}{\phi_{0}\Phi_{0}r}-\frac{2m}{\phi_{0}(2\omega+3\Phi_{0})r}+\frac{4\omega m}{3\phi_{0}\Phi_{0}(2\omega+3\Phi_{0})}\frac{e^{-m_{s}r}}{r}]\delta_{ij}.\label{gij}
\end{equation}
\begin{equation}
\phi=\phi_{0}[1+\frac{2m}{\phi_{0}(2\omega+3\Phi_{0})}\frac{1}{r}],\label{phi-variable}
\end{equation}
\begin{equation}
\Phi=\Phi_{0}[1+\frac{4\omega m}{3\phi_{0}\Phi_{0}(2\omega+3\Phi_{0})}\frac{e^{-m_{s}r}}{r}].\label{Phi-variable}
\end{equation}
Here the expression of the BD scalar field $\phi_{0}$ in the GBD theory can be gained as follows
 \begin{equation}
\phi_{0}=\frac{6\omega+12\Phi_{0}-2\omega e^{-m r}}{3\Phi_{0}(2\omega+3\Phi_{0})},\label{phi0}
\end{equation}
by comparing the metric component $g_{00}$ in the GBD  with the weak-field GR or Newton potential of a point mass.

\subsection{$\text{Discussions on PPN parameter in GBD theory}$}

 A gravity theory alternative to GR should be tested by the well-founded experimental results. As well known, the observational results can be directly applied to constrain the value of the parametrized post-Newtonian (PPN) parameter  $\gamma$. In addition, several debates can also be found in the $f(R)$ modified gravity theories. For example,   the value of PPN parameter  in $f(R)$ theories can be calculated to give $\gamma=\frac{1}{2}$  by using the   weak-field approximation method \cite{fr-review2} or other methods \cite{fr-gamma1,fr-gamma2,fr-gamma3}, which is a gross violation of the experimental bound $|\gamma-1|<2.3*10^{-5}$ \cite{bound-omega-gamma}. Or Ref. \cite{equal-fr-bd1}  originally claimed  that all $f(R)$ theories should be ruled out according to the fact that metric $f(R)$ gravity is equivalent to an  $\omega=0$  BD theory, since this theoretical predict value contradicts with the constraint: $ |\omega| > 40000$ deduced by the observational constraint on  $\gamma$ \cite{bound-omega-gamma}. The discussions on the equivalence between $f(R)$  and scalar-tensor gravity theories can be found in Refs. \cite{fr-review2,equal-fr-bd1,equal-fr-bd2,equal-fr-bd3,equal-fr-bd4}. The solutions to above contradictions were usually considered  by the following aspects:  the scalar is explained to be short-ranged (however, it cannot work as a model for late-time acceleration of universe \cite{fr-review2}), or there is even the possibility that the effective mass of the scalar field itself is actually scale-dependent \cite{fr-review2}, i.e. the so-called  chameleon mechanism----where the scalar has a large effective mass at terrestrial and Solar System scales,  while being effectively light at cosmological scales. So, exploring the solution to the problem of the PPN  parameter is valuable. Here we discuss the theoretical  value of the PPN parameter $\gamma$ in  the GBD theory and  compare its value with the observation.

 The concrete form of the PPN  parameter $\gamma$  in the GBD theory can be derived as follows
\begin{equation}
\gamma=\frac{h_{ii}}{h_{00}}=\frac{3\omega+3\Phi_{0}+\omega e^{-m_{s}r}}{3\omega+6\Phi_{0}-\omega e^{-m_{s}r}}.\label{gamma}
\end{equation}
 From Eq. (\ref{gamma}), one can see the dependence of the PPN parameter $\gamma$ with respect to model parameters: $\omega$, $\Phi_{0}$ and $m_{s}$. For  case of the heavy-mass scalar field (i.e. $m_{s}\gg 1$ or $e^{-m_{s}r}\rightarrow 0$), $\gamma\sim 1$ requires $\omega\gg \Phi_{0}$; For case of the light-mass scalar field  (i.e. $m_{s}\ll 1$ or $e^{-m_{s}r}\rightarrow 1$), $\gamma\sim 1$ requires $\omega= \frac{3\Phi_{0}}{2}$.  The different  mass $m_{s}$ of scalar field requires the different relations between the model parameters $\omega$ and $\Phi_{0}$. It is shown that the problem about the $\gamma$ value in the $f(R)$ theories is not existence in the GBD theory, even if we do not introduce  the chameleon mechanism in the latter theory. Also, Eq. (\ref{gamma}) can be seen as a viable condition for the GBD models.

\section{$\text{Gravitational waves physics in the GBD theory}$}
Studies on the GWs  in the gravity theory are very important.  The linearized framework from the weak-field approximation method provides a natural way to study the GWs. Lots of references paid attention to the studies on the GWs in the different aspects \cite{GW-other1,GW-other2,GW-other3,GW-other4,GW-other5,GW-other6,GW-other7,GW-other8}. In this paper, we are interested in the vacuum GWs of the GBD theory.  Considering an infinitesimal coordinate transformation, $x^{\mu}\rightarrow x^{'\mu}= x^{\mu}+\xi^{\mu}$, we obtain
\begin{equation}
h^{'}_{\mu\nu}= h_{\mu\nu}-\partial_{\mu}\xi_{\nu}-\partial_{\nu}\xi_{\mu},\label{hmunu-T}
\end{equation}
\begin{equation}
h^{'}= h-2\partial_{\sigma}\xi^{\sigma},\label{h-T}
\end{equation}
\begin{equation}
\theta^{'}_{\mu\nu}=h^{'}_{\mu\nu}-\frac{1}{2}\eta_{\mu\nu}h^{'}-\eta_{\mu\nu}\frac{\varphi^{'}}{\phi_{0}}+\eta_{\mu\nu}h_{f}^{'}
=\theta_{\mu\nu}+\eta_{\mu\nu}\partial_{\sigma}\xi^{\sigma}-\partial_{\mu}\xi_{\nu}-\partial_{\nu}\xi_{\mu},\label{thetamunu-T}
\end{equation}
\begin{equation}
\theta^{'}=\theta+2\partial_{\sigma}\xi^{\sigma},\label{theta-T}
\end{equation}
here $\xi^{\mu}$ is an arbitrary infinitesimal vector field with $|\xi^{\mu}|\ll 1$. If we choose $\xi^{\mu}$ to satisfy $\partial^{\mu}\theta_{\mu\nu}=\Box_{\eta}\xi^{\nu}$, then we again  preserve and get the gauge condition (\ref{Lorentz-gauge})
\begin{equation}
\partial^{\mu}\theta^{'}_{\mu\nu}=0.\label{Lorentz-GC}
\end{equation}
Thus, in the vacuum we solve wave Eq. (\ref{eq-box-theta}) to get
\begin{equation}
\theta_{\mu\nu}=A_{\mu\nu}(\vec{p})\exp(i k_{\alpha}x^{\alpha}).\label{solution-theta}
\end{equation}
Where $k_{\alpha}$  denotes the four-wavevector, and it is a null vector with $\eta_{\mu\nu}k^{\mu}k^{\nu}=0$. For GWs that propagate along the $z$-direction, $k^{\alpha}=\varpi(1,0,0,1)$ with $\varpi$ the angular frequency. From the  gauge condition  (\ref{Lorentz-gauge}), we can see that  the amplitude tensor  $A_{\mu\nu}$ is orthogonal to the direction of propagation of the waves $k^{\mu}A_{\mu\nu}=0$, which implies that the gauge freedom  can not be fixed completely. If $\xi^{\sigma}$ satisfies the equation $\theta^{'}=2\partial_{\sigma}\xi^{\sigma}$,  we then have $\theta=0$. To uniquely specify the perturbation, we have to search for four additional constraints on $A_{\mu\nu}$. Let us consider an observer detecting the gravitational radiation with describing by a unit timelike vector $u^{\alpha}=(1,0,0,0)$.  We can impose constraints $A_{0\nu}=0$, and obtain the components of the metric perturbation as follows
\begin{equation}
A_{\mu\nu}=\left(
\begin{array}{llll}
0 & 0& 0& 0 \\
0 & A_{+}& A_{\times}& 0 \\
0 & A_{\times}& -A_{+}& 0 \\
0 & 0& 0& 0
\end{array}
\right),\label{A-matrix}
\end{equation}
where $A_{+}$ and $ A_{\times}$  represent  the amplitudes of the two  independent polarization states of propagating gravitational radiation, just like electromagnetic waves. Obviously, in the above system of reference there are four non-zero components for the GBD gravity theory and they have the relations: $A_{11}=-A_{22}=A_{+}$ and $A_{12}=-A_{21}=A_{\times}$.  Eqs. (\ref{solution-theta}) and (\ref{A-matrix}) describe the gravitational radiation as appeared  in GR, while with the different expression of  $\theta_{\mu\nu}$.

 We can furthermore gain  the solutions of  the perturbation components $h_{\mu\nu}$  via  the relations: $h_{\mu\nu}=\theta_{\mu\nu}-\eta_{\mu\nu}\frac{\theta}{2}+\eta_{\mu\nu}h_{f}-\eta_{\mu\nu}\frac{\varphi}{\phi_{0}}$.  Putting the expression $h_{\mu\nu}$ in $g_{\mu\nu}=\eta_{\mu\nu}+h_{\mu\nu}$, the perturbed line element, due to the passing of a GW, can be described by
\begin{equation}
ds^{2}=(-1-h_{f}+\frac{\varphi}{\phi_{0}})dt^{2}+(1+A_{+}+h_{f}-\frac{\varphi}{\phi_{0}})dx^{2}+(1-A_{+}+h_{f}-\frac{\varphi}{\phi_{0}})dy^{2}
+(1+h_{f}-\frac{\varphi}{\phi_{0}})dz^{2}+2A_{\times}dxdy.\label{line-element}
\end{equation}
 Here  plane-wave solutions of the massless BD-field perturbation $\varphi$ and the massive geometry-field perturbation $h_{f}$ can be  given  by solving the wave equations (\ref{eq-box-varphi},\ref{eq-box-Phi}) with $T=0$  in the vacuum,
\begin{equation}
\varphi=a(\vec{p})\exp(i p_{\alpha}x^{\alpha}),\label{solution-varphi}
\end{equation}
\begin{equation}
h_{f}=b(\vec{p})\exp(i q_{\alpha}x^{\alpha}).\label{solution-hf}
\end{equation}
Obviously, the function $f_{R}$  of the Ricci scalar and the BD scalar field generate the new polarizations for GWs which are not present in the standard GR.

\section{$\text{Conclusion}$}
Several observational and theoretical  motivations require us to investigate the  modified or alternative theories of GR.  Lots of modified gravity theories have been proposed and widely studied, especially two simple modified gravity of GR: the $f(R)$ theory and the  Brans-Dicke  theory. In the BD  theory,  the BD scalar field can be introduced naturally by considering a time-variable Newton gravity constant. Many extended versions of the BD theory have been explored and developed.  In this paper, we  explore a modified Brans-Dicke theory by generalizing the Ricci scalar $R$  in the original BD action to an arbitrary function $f(R)$.  Comparing with other theories, it can be found   in Ref. \cite{GBD} that the GBD theory have some interesting properties or solve some problems existing in other theories. (1)  One knows that in the couple scalar-fields quintom model, it is required to include both the canonical quintessence field and the non-canonical phantom field  in order to make the state parameter to cross over $w=-1$, while several fundamental problems are associated with phantom field, such as the problem of negative kinetic term and the fine-tuning problem, etc. While, in the GBD model, the state parameter of geometrical dark energy can cross over the phantom boundary $w=-1$ as achieved in the quintom model,  without bearing the problems existing in the quintom model. (2) It is well known that the $f(R)$ theory are equivalent to the BD theory with a potential (abbreviated as BDV) for taking a specific value of the BD parameter $\omega=0$, where the specific choice: $\omega=0$ for the BD parameter is quite exceptional, and it is hard to understand the corresponding absence of the kinetic term for the field. However, for the GBD theory, it is similar to the couple scalar-fields model, and both fields in the GBD own the non-disappeared dynamical effect.  In addition, the GBD theory tends to investigate the physics from the viewpoint of geometry, while the BDV or the two scalar-fields quintom model tends to solve physical problems from the viewpoint of matter. It is possible that several special characteristics of scalar fields could be revealed through studies of geometrical gravity in the GBD.

 In this paper, we continue to investigate the GBD theory.  Using the method of the weak-field approximation, we explore the linearized physics in the   GBD theory. The linearized equations of the gravitational field and two scalar fields  are given.  We investigate their solutions in the linearized theory for a point mass. It is shown that the problem of the $\gamma$ value in the $f(R)$ theory could be solved in the GBD theory, where the theoretical $\gamma$ value in the GBD theory can be consistent with the observational results.  At last, we study the gravitational waves physics in the vacuum for the GBD theory. It is found that the gravitational radiation in the GBD theory has new freedoms beyond the two standard modes in the general relativity theory.

\textbf{\ Acknowledgments }
We thank the anonymous referee for his/her very instructive comments, which improve our paper greatly.  The research work is supported by   the National Natural Science Foundation of China (11645003,11705079,11575075,11475143).

\end{document}